# On the Impact of LTE-U on Wi-Fi Performance

Alireza Babaei, Jennifer Andreoli-Fang, Belal Hamzeh
Cable Television Laboratories
Louisville, CO, USA

*Abstract*—With the exponential growth in mobile data traffic taking place currently and projected into the future, mobile operators need cost effective ways to manage the load of their networks. Traditionally, this has been achieved by offloading mobile traffic onto Wi-Fi networks due to their low cost and ubiquitous deployment. Recently, LTE operating in the unlicensed spectrum has drawn significant interests from mobile operators due to the availability of the unlicensed spectrum. However, the deployment of LTE networks in the unlicensed band poses significant challenges to the performance of current and future Wi-Fi networks. We discuss the LTE and Wi-Fi coexistence challenges and present analysis on performance degradation of the Wi-Fi networks at the presence of LTE.

## I. Introduction

The exponential growth of mobile data traffic is driving mobile network operators (MNOs) to look into various cost effective solutions to meet the continuously increasing demand and offload traffic from the licensed spectrum. The low cost of Wi-Fi access points, the pervasiveness of Wi-Fi in mobile devices and the availability of unlicensed spectrum has made Wi-Fi the technology of choice for data offload. Nonetheless, the integration of Wi-Fi into the 3GPP core network remains complex despite the availability of four separate standardized methods dating back to 3GPP Release 6 [1]. Despite the numerous options, none of them were found to be satisfactory by the MNOs and thus no wide deployments of the solutions are seen.

Most recently, the 3GPP is considering extending the use of LTE into the unlicensed spectrum as another means to enable traffic offload. This new approach is dubbed LTE Unlicensed (LTE-U). Compared to Wi-Fi, LTE-U offers MNOs a way to offload traffic onto the unlicensed spectrum with a technology that seamlessly integrates into their existing LTE evolved packet core (EPC) architecture. A single eNodeB can support LTE and LTE-U for seamless integration to the MNO network. Furthermore, LTE-U promises higher throughout and spectral efficiency than Wi-Fi, with estimates ranging from 2x to 5x improvement over Wi-Fi [2, 3].

Three modes have been proposed for LTE-U, distinguished by the supplementary and control channel configurations as shown in Fig. 1:

- Supplemental Downlink (SDL): In this mode, the unlicensed band is used to solely carry data traffic in the downlink direction, while the uplink and control channel remain in the licensed spectrum. This option has been proposed in the 3GPP.

- Carrier Aggregation TD-LTE: In this mode, the unlicensed band is used as an auxiliary TDD channel capable of carrying data traffic in the uplink and the downlink directions while the control channel remains in the licensed spectrum. This option has also been proposed in the 3GPP.

- Standalone LTE-U: In this mode, the data and the control channels of LTE-U operate in the unlicensed spectrum; thus there is no dependence on licensed spectrum availability to support LTE-U operations. This option has not been discussed in the 3GPP, but provides a option for operators that do not currently own spectrum to benefit from LTE-U capabilities.

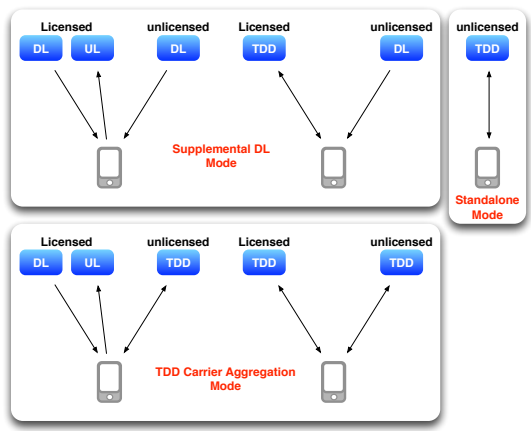

Fig. 1. LTE-U alternatives

In this paper, we consider the potential impact of LTE-U on Wi-Fi networks for the two configurations being proposed in the 3GPP. We begin with a brief review of the lower layers of LTE and Wi-Fi protocols in Section III, followed by an analysis of the LTE "quiet period" in Section IV. We then present a probabilistic framework to determine the likelihood of Wi-Fi transmission during the LTE quiet period. Numerical results are presented in Section V.

## II. Prior Works

The problem of Wi-Fi and LTE coexistence and the potential impact of one network over the other have recently been studied and simulation results have been presented in a handful of research and industry publications.

In [4], a paper published by Nokia Research, a simulator-based system-level analysis has been performed to assess the performance of LTE and Wi-Fi networks coexisting in an office environment. Single-floor and multi-floor office environments with different assumptions on the density of Wi-Fi and LTE nodes have been considered in the simulation. Although the simulation model, the assumptions on Wi-Fi and LTE system parameters and deployment environment can be improved, the results presented in [4] validate our analysis presented in this paper: channel sharing between Wi-Fi and LTE networks is significantly unfair for the Wi-Fi network.

In [3], a whitepaper published by Qualcomm, LTE-U is described as a better neighbor to Wi-Fi than Wi-Fi to itself. It is also claimed that LTE-U provides operators substantial improvements in data throughput without any impact to Wi-Fi users when Qualcomm's proprietary coexistence mechanisms are applied. While these claims are derived from simulations, the simulation models used are not available publically.

In [2], Huawei provides the result of their simulation on spectrum efficiency comparison between Wi-Fi and LTE in a sparse deployment scenario. It states that the simulation includes coexistence updates to LTE-U to accommodate Wi-Fi, but does not provide sufficient detail on the effectiveness of the coexistence features. The trends of interference based on traffic load appear credible, if LTE-U to LTE-U coordination is achieved or interference avoidance is deployed.

### III. A COMPARISON OF WI-FI AND LTE LOWER LAYERS

Although both Wi-Fi and LTE PHY layers are based on the OFDM technology, their transmissions are not orthogonal due to different subcarrier spacing and lack of synchronization.

TABLE I. MAC LAYER COMPARISONS

| | *LTE* | *Wi-Fi* |
|---|---|---|
| Multiple access | Multiple users served simultaneously, occupying different frequencies in channel | Absent of MU-MIMO, only 1 user is served at a time, takes up entire channel spectrum |
| Channel usage | Frames are contiguous, so channels are approximately "always on" | Channel is occupied only when packets needs to be transmitted |
| Channel access | Centralized scheduling on DL and UL. LTE does not contend, it simply transmits | Distributed Coordination Function (DCF), contention-based[1] |
| Collision avoidance | None, b/c channel access are centrally scheduled | CSMA/CA + RTS/CTS (In principle, "sense before transmit") |
| Co-existence | Has not had the need to be able to coexist with other technologies | Already coexists well with other technologies in unlicensed band, although with no common fairness mechanism |

Table I shows that LTE MAC may be more efficient at spectrum usage compared to Wi-Fi MAC, specially when large number of users access the medium. This is primarily due to the centralized scheduling nature of the LTE protocol at the eNodeB. LTE will fill the airtime when the traffic load permits. The maximum sector capacity is independent of the number of UEs being served by the LTE eNodeB. On the other hand, as the number of users increases in the Wi-Fi network, the performance of CSMA/CA and channel utilization degrades due to the increased probability of collision [5].

### IV. THE COEXISTENCE CHALLENGE

LTE-U poses significant coexistence challenges for Wi-Fi networks due to the inherent differences between channel usage and access procedures used by each technology. Wi-Fi is designed to coexist with other technologies through channel sensing and random backoff. On the other hand, LTE is designed with the assumption that one operator has exclusive control of a given spectrum; LTE traffic channels are designed to continuously transmit with minimum time gap even in the abscense of data traffic. Consequently, Wi-Fi users will have little chance to sense a clear channel and deem it suitable for transmission.

LTE is an almost continuously transmitting protocol. In order for Wi-Fi users to transmit, they need to wait for a "quiet" period when LTE is not transmitting. Even when there is no data traffic present on the air interface, LTE periodically transmits a variety of control and Reference Signals. How long LTE remains truly "quiet" depends on the periodicity of these signals. We examine the control signals next.

#### A. LTE "Quiet Period"

*1) LTE FDD System*

As shown in Fig. 1, LTE-U can be achieved via several different access modes, depending on which link is carried on the unlicensed spectrum. One realization of LTE-U is to carry one or more DL carrier(s) of a LTE FDD network on the unlicensed spectrum. The periodicity of the DL control and Reference Signals will dictate whether and when Wi-Fi may be able to leverage these quiet periods and be able to transmit.

Fig. 2 shows a pair of LTE DL Resource Blocks (RBs) with Physical Downlink Control Channel (PDCCH) and Reference Signals [2]. The PDCCH carries UL and DL scheduling assignments, among other vital control information. The PDCCH occurs at the start of every subframe, or every 1 ms, taking between 1 and 3 OFDM symbols. The Reference Signals are present regardless of whether DL data transmissions are present, and are used for channel estimation for coherent detection. The Reference Signals are transmitted in every DL subframe at fixed locations, spanning the entire DL bandwidth.

LTE transmits other periodic signals such as Primary and Secondary Synchronization Signals on the DL. The periodicity of these signals are much longer than the Reference Signals.

The periodicity of the Reference Signals dominates the duration of the "quiet" periods on the LTE DL, with a maximum of 3 symbol periods, or approximately 215 µs.

---

[1] Although Point Coordination Function is also defined in Wi-Fi, it is not widely implemented, and therefore not discussed in this paper.

[2] Only standard slot configuration is considered, i.e., normal cyclic prefix with 7 symbols per slot. No MIMO configuration is considered.

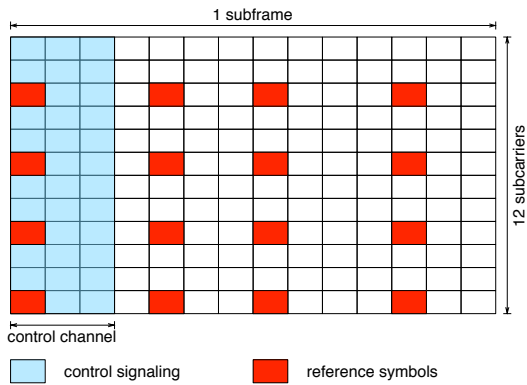

Fig. 2. LTE DL control and reference signals

*2) TD-LTE "Quiet Period"*

Another realization of LTE-U shown in Fig. 1 is to carry both the UL and DL traffic of a TD-LTE network on the unlicensed spectrum. In TD-LTE, seven UL/DL configurations are defined to allow for the adaptation of different UL-DL traffic profiles by assigning more or less subframes within a frame for UL or DL data transmission.

To enable fair access to the channel in the unlicensed spectrum, LTE-U using TD-LTE network may be designed to intentionally not schedule data transmission for X subframes during the period of every Y total subframes. For example, UL/DL configurations 0, 3, and 6 all show that a maximum of 3 UL subframes (or 3 ms) are scheduled together, and therefore can be intentionally muted by the eNB. This duty cycle approach to coexistence allows LTE-U to maintain the efficiencies it enjoys due to the scheduled nature of the LTE air interface while providing WiFi APs opportunities to access the channel.

### B. How does LTE Quiet Period Compare to Wi-Fi?

For LTE-U, the maximum quiet period is

- 3 symbols, or approximately 215 µs, on the DL of a LTE FDD network
- Up to 3 subframes, or 3 ms, on a TD-LTE network

Wi-Fi AP and devices need to back off for a random period of time prior to transmission which can potentially occur outside the window of the LTE quiet period. When a transmission does occur, the burst length for a 1518 byte frame is approximately between 110 µs and 1.8 ms, depending on the modulation and coding used.

Unless the LTE-U traffic channels are designed differently than LTE traffic channels in licensed spectrum, LTE-U will apply continuous traffic to devices in a periodic fashion. LTE-U will present signficant challenges to Wi-Fi throughput and delay performances by maintaining control of a large share of the airtime.

### C. Probability of Wi-Fi Channel Access

In Section IV.A, we derived the maximum "quiet" period for SDL and TD-LTE modes of LTE-U. In this Section, our goal is to obtain the probability of Wi-Fi channel access, i.e., the probability that Wi-Fi backoff delay is less than the LTE-U quiet periods[3]. By Wi-Fi backoff delay, we mean the time elapsed since Wi-Fi starts its backoff process until the packet is successfully transmitted. Let us define $d$ as the random variable denoting backoff delay and $L$ as the length of LTE-U quiet period. With the above notations, the probability of Wi-Fi channel access within an LTE-U quiet period is $\Pr\{d < L\}$.

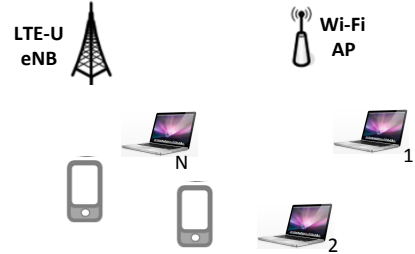

Fig. 3. LTE-U network coexisting with a Wi-Fi network with N stations

We consider a Wi-Fi network with $N$ stations coexisting with an LTE-U network as shown in Fig. 3. We assume that the $N$ stations follow the DCF and backoff rules to access the channel. The $N$ stations in the Wi-Fi network contend to access the wireless medium and the collisions of their transmitted packets increases the backoff delay. In Fig. 4, we show different components of backoff delay.

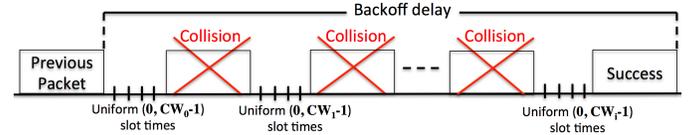

Fig. 4. Backoff delay components

For a Wi-Fi network consisting of $N$ stations, the probability distribution of backoff delay has been studied in [6]. Here, we use the analysis of [6] as a basis to calculate the probability of an arbitrary Wi-Fi client having backoff delay less than the LTE-U quiet period.

The amount of backoff delay depends on the number of collisions before the successful transmission of a packet. Using the total probability theorem, we have

$$\Pr\{d < L\} = \sum_{i=0}^{R} \Pr\{d < L | i \text{ collisions}\} \Pr\{i \text{ collisions}\} \quad (1)$$

where $R$ is the retry limit, i.e., the maximum number of collisions before the packet is discarded. Define a slot time as the time duration between two consecutive backoff decrements. Note that, a slot time can be empty (which lasts as long as the system-defined time slot), or may contain the transmission of one or more stations (in which case a freeze on backoff timer decrement happens). Expanding (1), we have

$$\Pr\{d < L\} = \sum_{i=0}^{R} \sum_{j=0}^{W_i} \Pr\{d < L | i \text{ collisions}, j \text{ slots}\} \Pr\{j \text{ slots} | i \text{ collisions}\} \Pr\{i \text{ collisions}\} \quad (2)$$

where $W_i = \sum_{k=0}^{i} CW_k - 1$, $CW_k = \min(2^k CW_{min}, CW_{max})$ is the maximum contention window size after $k$ collisions, and $CW_{min}$ and $CW_{max}$ are the minimum and maximum contention window size as defined in 802.11 standard. Note that for $CW_k < CW_{max}$, we have $CW_k = 2CW_{k-1}$, i.e., contention window size doubles after each collision. The three components in the above summation (i.e., $\Pr\{d < L|i \text{ collisions}, j \text{ slots}\}$, $\Pr\{j \text{ slots}| i \text{ collisions}\}$ and $\Pr\{i \text{ collisions}\}$) are obtained in [6]. Specifically, to calculate $\Pr\{j \text{ slots}| i \text{ collisions}\}$, we note that with $i$ collisions, the number slot times is sum of $i+1$ uniformly distributed random variables, i.e. $\sum_{k=0}^{i} \text{unif}(0, CW_k - 1)$, where $\text{unif}(0, CW_k - 1)$ is a uniform random variable with lower and upper limits of 0 and $CW_k - 1$ (See Figure 4). Consequently,

$$\Pr\{j \text{ slots}| i \text{ collisions}\} = \Pr(\sum_{k=0}^{i} \text{unif}(0, CW_k - 1) = j) \quad (3)$$

---

[3] Note that, even if the LTE-U quiet period is long enough that the Wi-Fi user can access the channel, its transmission may be interfered by LTE-U users, seriously degrading its throughput. This is however outside the scope of this Section.

The probability mass function (PMF) of sum of $i+1$ uniform random variables can be found from the convolution of individual PMFs. Using the PMF of sum, the above probability can be found in closed-form. Also, we have

$$\Pr\{i \text{ collisions}\} = P_c^i P_s, \quad (4)$$

where $P_s$ is the probability of successful transmission (i.e., the probability that only 1 out of $N$ stations transmit) and $P_c$ is the probability of collision (i.e., the probability that more than 1 station transmits). Denoting $\tau$ as the probability that a station transmits at an arbitrary slot time, we have

$$P_s = (1 - \tau)^{N-1},$$
$$P_c = 1 - (1 - \tau)^{N-1}.$$

The value of $\tau$ is obtained in closed-form in [7, 8].

The last component, i.e., $\Pr\{d < L|i \text{ collisions}, j \text{ slots}\}$ is also found in [6]. The analysis in [6], assumes that distribution of backoff delay given $i$ collisions and $j$ slots follows a Gaussian distribution. Denoting the mean and variance of the Gaussian random variabl as $m_{ij}$ and $\sigma_{ij}$, we have

$$\Pr\{d < L| i \text{ collisions}, j \text{ slots}\} = \begin{cases} 0.5 + 0.5 \text{ erf}\left(\frac{D - m_{ij}}{\sqrt{2}\sigma_{ij}}\right) & \frac{D - m_{ij}}{\sqrt{2}\sigma_{ij}} \geq 0 \\ 0.5 \text{ erf}\left(-\frac{D - m_{ij}}{\sqrt{2}\sigma_{ij}}\right) & \frac{D - m_{ij}}{\sqrt{2}\sigma_{ij}} < 0 \end{cases} \quad (5)$$

The values of $m_{ij}$ and $\sigma_{ij}$ depend on duration of interframe spacing (SIFS, DIFS and EIFS), packet size, ACK size, MAC overhead, physical layer convergence protocol (PLCP) preamble and header transmission time, and duration of an empty slot time, among other DCF parameters (See [6]). Particularly, for smaller the packet size, mean and variance of the Gaussian random variable will be smaller.

Substituting (3), (4) and (5) in (2), we can find the cumulative distribution function (CDF) of backoff delay (i.e., $\Pr\{d < L\}$). The statistical mean of backoff delay can be found from its CDF as follows [9]:

$$E\{d\} = \int_0^\infty (1 - \Pr\{d < L\}) \, dL. \quad (6)$$

Statistical mean of backoff delay can be obtained numerically from (6) after finding the CDF of $d$ from (1). In next Section, we use the DCF parameters used in 802.11n to calculate the probability of backoff delay.

## V. PERFORMANCE EVALUATION

In this Section, we present numerical results based on the analysis performed in Section IV to evaluate the coexistence challenges between LTE-U and Wi-Fi networks.

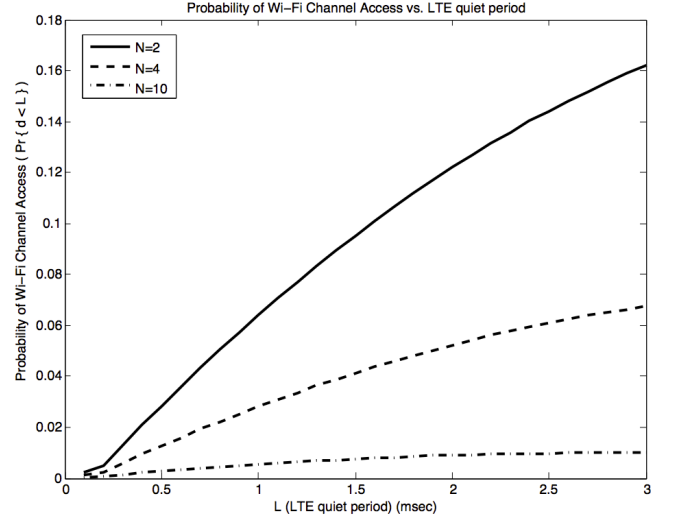

Fig. 5. Probability of Wi-Fi channel access vs. LTE quiet period (Packet size=1500 Bytes)

In Fig. 5, assuming a fixed Wi-Fi packet size of 1500 bytes, the probability of Wi-Fi channel access is shown versus the LTE quiet period. As we discussed in Section IV, the maximum quiet period that can be created by muting UL subframes in the TD-LTE mode is 3 $msec$. Fig. 5 shows that, even when the number of Wi-Fi stations is as low as $N = 2$ (i.e., very light contention) and the LTE-U quiet period is as high as 3 $msec$, the probability that backoff delay is smaller than LTE-U quiet period is very small (about 0.16). This probability is even smaller when the number of Wi-Fi users increases. In other words, the probability that a Wi-Fi station can have the chance to access the medium in the presence of a LTE-U network is very small (about 16% in the best case).

In Figure 6, assuming 4 Wi-Fi users (i.e., $N = 4$), the same probabilities are found 3 different packet sizes of 500 bytes, 1000 bytes and 1500 bytes. As described in Section IV, for smaller packet size, the conditional probabilities found in (5) will be larger and as a result, the probability of Wi-Fi channel acccess will also increase.

Figure 7 shows the statistical average of backoff delay versus the number of Wi-Fi stations. Mean backoff delay is obtained using equation (6) in Section V. The results indicate that even when the number of Wi-Fi users is as low as 2 and with the Wi-Fi packet size is as small as 500 bytes, mean backoff delay (about 4 $msec$) is larger than the maximum LTE quiet period (3 $msec$). Increasing the packet size or number of Wi-Fi stations increases the mean backoff delay as expected.

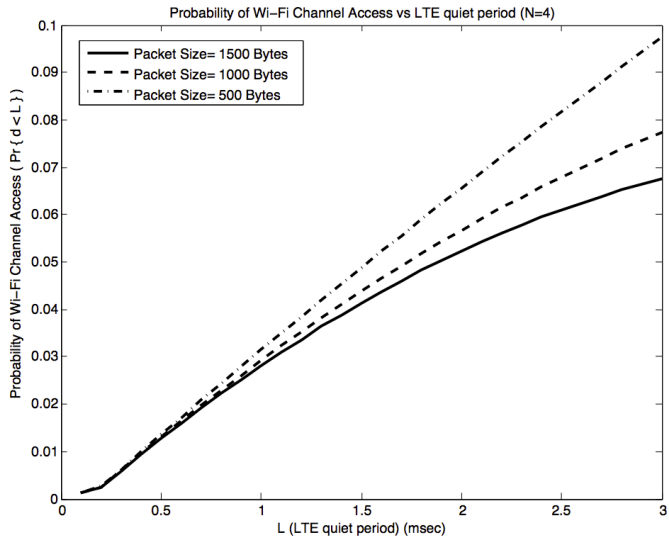

Fig. 6. Probability of Wi-Fi channel access vs. LTE quiet period ($N = 4$)

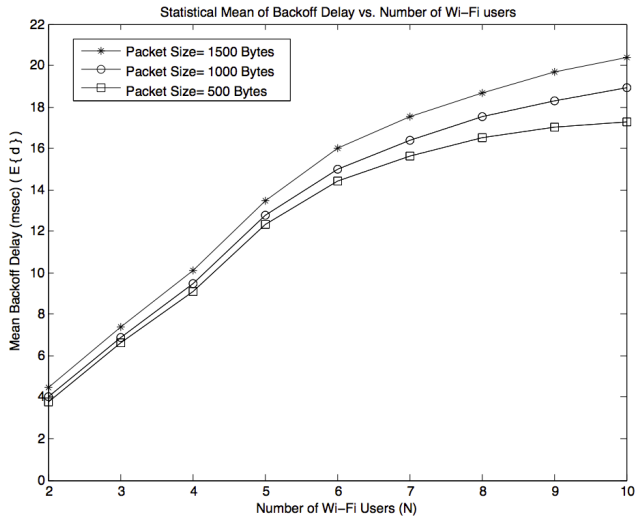

Fig. 7. Statistical Mean of Backoff Delay vs. $N$

## VI. Conclusions

Our probabilistic and numerical analyses show that when Wi-Fi and LTE networks operate together in the unlicensed band without modifications to existing protocols, Wi-Fi transmissions are significantly affected by the presence of LTE transmissions. Specifically, given the two potential modes of operations currently proposed for LTE-U in the unlicensed spectrum, the amount of "quiet" period presented by the LTE protocol for Wi-Fi users is too short to allow access to the channel. As a result, Wi-Fi is at risk of spending a significant amount of time in the "listening" mode when LTE transmission is present in the same channel.

Our results indicate that much work needs to be done to achieve a "fair" coexistence mechanism. LTE MAC layer will need to be redesigned if Wi-Fi is to be afforded a useful portion of the unlicensed spectrum. But how best to design coexistence into LTE-U without substantially degrading the data throughput efficiency of LTE-U remains an open question.

Ideally, coexistence requirements and solutions should provide a level playing field for each network and technology. Airtime fairness and data throughput efficiency are both important considerations, although it may be difficult to achieve both in the case of coexistence of LTE and Wi-Fi. On the one hand, one could argue that coexistence mechanisms should ideally provide each network an equal opportunity for airtime fairness. Specifically, each network needs to be able to utilize equivalent portions of spectrum over time as traffic conditions meet or exceed the data throughput capacity of the air interface. This does not necessarily provide each device in the network the same average data rate, which is dependent upon a number of factors. Airtime fairness shares equivalent megahertz portions of spectrum equally among participants.

Regulatory requirements are designed to provide a certain level of airtime fairness, with arguable results towards fairness at the data throughput efficiency level. The U.S. and China do not mandate specific coexistence requirements for 5 GHz unlicensed spectrum. Europe, however, does mandate the coexistence requirements as summarized in [10].

On the other hand, coexistence mechanisms should also strive for data rate efficiency. But a range of coexistence techniques to help ensure airtime fairness may present costs to data rate efficiency. A significant portion of the LTE efficiency is due to the centralized and continuously scheduled nature of its air interface. If LTE-U were to be subject to the inefficiencies of the Wi-Fi's "listen before talk" procedures, it would lose some of the benefit of LTE's scheduled air interface.